# Temperature-dependent Structural Evolution of Ruddlesden–Popper Bilayer Nickelate $La_3Ni_2O_7$

Haozhe Wang[1], Haidong Zhou[2], Weiwei Xie[1]*

1. Department of Chemistry, Michigan State University, East Lansing, MI 48824, USA
2. Department of Physics and Astronomy, University of Tennessee, Knoxville, TN 37996, USA

* Email: xieweiwe@msu.edu

*Abstract*

A recent *J. Am. Chem. Soc.* Article (DOI: 10.1021/jacs.3c13094) details a pressure-temperature (*P-T*) phase diagram for the Ruddlesden–Popper bilayer nickelate $La_3Ni_2O_7$ (LNO-2222) using synchrotron X-ray diffraction. This study identifies a phase transition from *Amam* (#63) to *Fmmm* (#69) within the temperature range of 104 K to 120 K under initial pressure and attributes the *I*4/*mmm* (#139) space group to the structure responsible for the superconductivity of LNO-2222. Herein, we examine the temperature-dependent structural evolution of LNO-2222 single crystals at ambient pressure. Contrary to symmetry increase and the established *Amam-Fmmm* phase boundary, we observe an enhancement in the *Cmcm* reflections as temperature decreases. This work not only establishes a benchmark method for single crystal structure studies of LNO-2222 using laboratory X-rays, but also enhances the understanding of the complex crystallographic behavior of this system, contributing insights to further experimental and theoretical explorations.

Nickelates have emerged as promising candidates for high-temperature superconductivity, drawing parallel with the cuprates first discovered in the 1980s,[1] due to their analogous crystal and electronic structures.[2] Notably, superconductivity was previously identified in epitaxial thin films of reduced square-planar phases,[3-9] characterized by an ultra-low valence state of $Ni^{1+}$, isostructural to $Cu^{2+}$. Continuing this trajectory, a significant breakthrough was reported last year with the observation of superconductivity signatures in Ruddlesden–Popper bilayer nickelate $La_3Ni_2O_7$ (LNO-2222), which exhibited a $T_C$ up to 80 K within a pressure range of 14.0 GPa to 43.5 GPa.[10] This discovery was also marked by a structural transition from the ambient pressure *Amam* (#63) to the high-pressure *Fmmm* (#69) above 15.0 GPa, aligning with the onset of superconductivity. Further advancements have been achieved through *in situ* low-temperature high-pressure synchrotron X-ray diffraction (XRD), which has effectively mapped the pressure-temperature (*P-T*) structure phase diagram of LNO-2222, clarifying the phase boundaries among the *Amam*, *Fmmm*, and *I4/mmm* (#139) space group.[11]

Initially, this study caught our attention, because a phase transition from *Amam* to *Fmmm* was observed within the temperature range of 104 K to 120 K under initial pressure conditions.[11] However, this symmetry increase contradicts our previous observations in high-quality Sr-doped LNO-2222 single crystals synthesized under high pressure.[12] The absence of detailed crystal structure data targeting the *Amam* to *Fmmm* transition in the study prompted further investigation into the temperature effects on this specific symmetry change. Meanwhile, the structure determination in the referenced study[11] was conducted using powder XRD refinements, achieving a resolution limit of approximately 1.14 Å. While it is recognized that obtaining higher resolution and enhanced data quality under high pressure presents significant challenges, in the case of LNO-2222, precise determination of lattice centering and space groups critically depends on the observation of some weak peaks. Therefore, a more dedicated focus on studying the crystal structure, particularly through the use of single crystals, is essential for comprehensively resolving the complexities of this material.

On the other hand, increasing efforts have been directed toward theoretical investigations of the mechanism behind high-$T_C$ superconductivity in LNO-2222. The accuracy of these theoretical models might be compromised without precisely determined crystal structures. Notably, conflicting suggestions regarding the size of A-site cations and their role in stabilizing the superconducting phase at ambient pressure have been reported.[13,14]

We conducted temperature-dependent single crystal XRD experiments on LNO-2222, spanning a temperature range from 80 K to 400 K at ambient pressure. Our results indicate that with decreasing temperature, the tilt of octahedra in LNO-2222 is enhanced, and the *Cmcm* space group becomes increasingly favorable. This work aims to establish a benchmark method for single crystal structure studies of LNO-2222 using laboratory X-rays. Further investigations exploring the crystal structure and structure-property relationship in LNO-2222 will likely require much more high-quality data, utilizing advanced techniques such as synchrotron X-rays and neutrons.

Procedures of our single crystal XRD experiments and data analysis of LNO-2222 are detailed in **Scheme 1**. The crystals of LNO-2222 used in this study were selected from the same batch as previously reported[15] and grown using the floating zone method under 100% $O_2$ at a pressure of 14–15 bar at the University of Tennessee. A single crystal with dimensions of 0.069 × 0.045 × 0.014 $mm^3$ was picked up, mounted on a nylon loop with paratone oil, and measured using an XtalLAB Synergy, Dualflex, Hypix single crystal X-ray diffractometer equipped with an Oxford Cryosystems 800 low-temperature device. Prior to initiating temperature-varied data collection, the quality of the crystal was examined at room temperature, confirming its bilayer stacking. The temperature protocol commenced with cooling the sample to 80(2) K, followed by sequential heating to 400(2) K in increments of 40 K. A 10-minute stabilization period was allowed between each temperature scan, with continuous monitoring and adjustment of crystal centering as needed throughout the process. Data acquisition was performed using $\omega$ scans with Mo $K_\alpha$ radiation ($\lambda$ = 0.71073 Å, micro-focus sealed X-ray tube, 50 kV, 1 mA). The measurement strategy, including the total number of runs and images, was determined using the strategy calculation feature in CrysAlisPro software (version 1.171.43.104a, Rigaku OD, 2023), which was established at 80(2) K and consistently applied across all temperature steps. Data reduction induced correction for Lorentz polarization. Numerical absorption correction based on Gaussian integration over a multifaceted crystal model. Empirical absorption correction was applied using spherical harmonics implemented in SCALE3 ABSPACK scaling algorithm. Structure solution and refinement were conducted using the Bruker SHELXTL Software Package.[16,17]

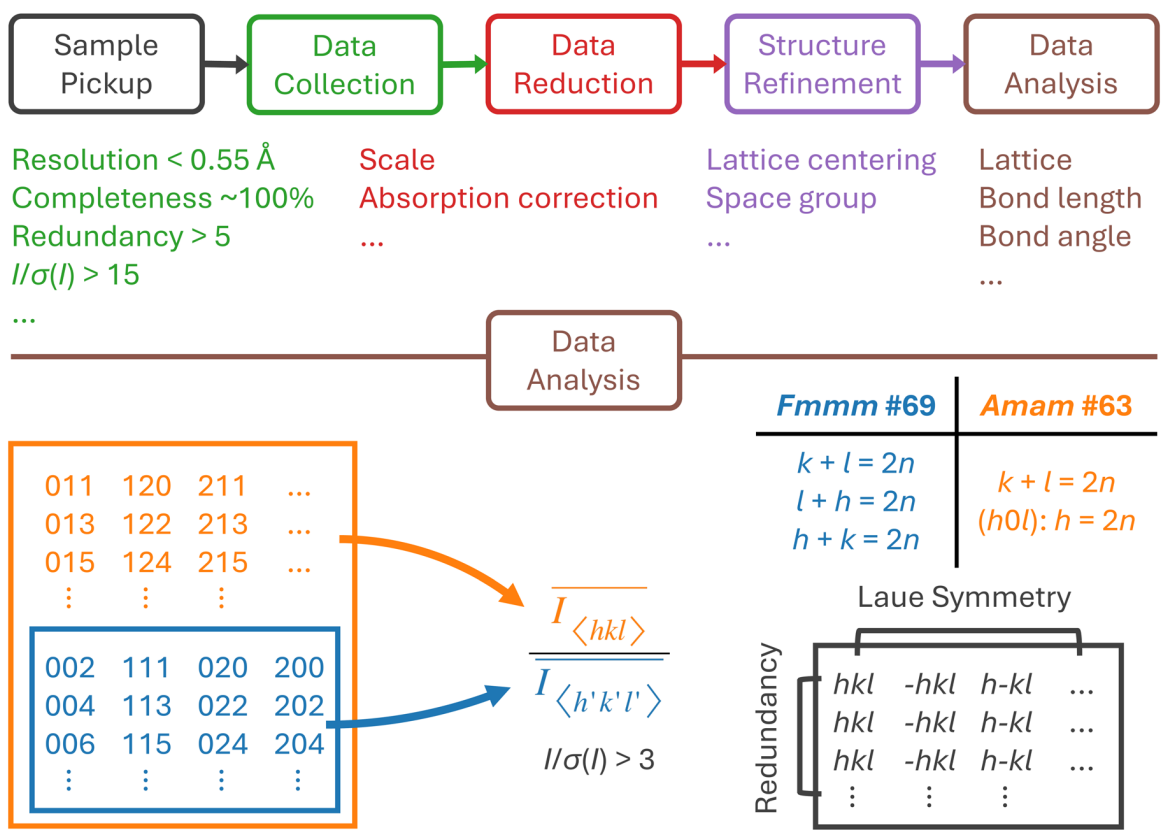

**Scheme 1.** Procedure of our single crystal XRD experiments and data analysis of LNO-2222.

The initial determination of the crystal structure of LNO-2222, classified in the *F*-centered orthorhombic *Fmmm* space group, was performed using powder XRD and Rietveld refinement on polycrystalline samples.[18] Recognizing the substantial uncertainty in determining oxygen coordinates, neutron powder diffraction (NPD) was subsequently employed.[19,20] The results indicated that the *Fmmm* space group was inappropriate, as it failed to account for extra weak peaks, suggesting a lower symmetry. Consequently, a *C*-centered orthorhombic lattice in the space group *Cmcm* was proposed.[19] Challenges related to the lattice centering, the impact of oxygen vacancies, and the coexistence of Ruddlesden–Popper bilayer and trilayer phases highlight the complexities in structure determination, underscoring the necessity for pure LNO-2222 single crystals. Recent advances in the high-pressure floating zone method, which allows for a 100% O$_2$ atmosphere with controllable gas pressure, have facilitated the growth of high-purity Ruddlesden–Popper nickelate single crystals, enabling precise structure determination using laboratory X-rays.[15,21,22] Furthermore, a previously unrecognized phase of $La_3Ni_2O_7$ with distinct layer stacking, LNO-1313, was first reported by Chen *et al.*[22] in the growth of LNO-2222, and supported by later studies.[15,23]

In the referenced low-temperature high-pressure study,[11] multiple instances of twins and severe texture development were reported during pressure increase across multiple runs, yet these claims lacked experimental evidence. If the sample quality was thoroughly examined prior to pressurization, it would be essential to document at which pressure and temperature conditions these multiple twins (or texture development) occurred in LNO-2222, their intrinsic nature, and their impacts on the results of structure determination. The results from powder XRD refinement results remain ambiguous, with "atomic positions optimized theoretically" mentioned but without additional details provided.

The presence of oxygen vacancies significantly influences the determination of crystal structure, often leading to a symmetry increase. For example, polycrystalline samples of $La_3Ni_2O_{6.92}$-2222 and $La_3Ni_2O_{6.94}$-2222 were determined to be *Fmmm* by powder XRD,[18,24] while $La_3Ni_2O_6$-2222 and $La_3Ni_2O_{6.35}$-2222 was identified as $I4/mmm$ by NPD.[25,26]

Concerning the tuning of LNO-2222 crystal structures through A-site doping, there are claims that using smaller atoms might induce a chemical precompression effect.[11] One argument presented is the comparison of the A/B atomic size ratio between tetragonal $Sr_3Ti_2O_7$ and LNO-2222. However, the lack of a referenced source for atomic size data undermines the credibility of this comparison. Whether using atomic size or ionic size is more appropriate remains unclear. According to the authors' logic, a rough calculation reveal that $Sr^{2+}/Ti^{4+}$ = 132/74.5 = 1.772, while $La^{3+}/(0.5*(Ni^{2+}+Ni^{3+}))$ = 117.2/(0.5*(83+70)) = 1.532.[27] Even considering $Ni^{2+/3+}$ as $Ni^{3+}$ exclusively, the upper boundary for $La^{3+}/Ni^{2+/3+}$ would be 117.2/70 = 1.674, still smaller than the $Sr^{2+}/Ti^{4+}$ ratio, suggesting an opposite conclusion. Furthermore, in the case of $La_{2-2x}Sr_{1+2x}Mn_2O_7$, detailed crystallographic and magnetic phase diagram provided by temperature-dependent neutron powder diffraction shows a tetragonal-to-orthorhombic phase transition, peaking at $x = 0.80$.[28] However, these findings were never related to the size of A-site in the original report. The persistence of the tetragonal $I4/mmm$ structure to 35 GPa in $LaSr_2Mn_2O_7$[29] ($x = 1.00$) contradicts the claims, as the size of $Sr^{2+}$ is larger than $La^{3+}$.

Our previous report[12] included a temperature-dependent crystal structure study of high-quality Sr-doped LNO-2222 single crystals (formula $La_{2.80(1)}Sr_{0.20(1)}Ni_2O_{6.95(1)}$, denoted as Sr-LNO-2222), obtained via high-pressure synthesis. A comparison of the out-of-plane Ni–O–Ni bond angles between undoped LNO-2222 and Sr-LNO-2222 reveals that Sr-LNO-2222 exhibits fewer octahedral tilts, which supports the hypothesis that the incorporation of larger A-site atoms contributes to a potential rise in symmetry.

**Figure 1** illustrates the crystal structure of LNO-2222 in the space groups *Cmcm*, *Fmmm*, and *I*4/*mmm*, providing a detailed view of their structural variations. Central to this discussion are the group-subgroup relationships that explain how LNO-2222 crystallizes in these specific space groups. In the *Cmcm* space group, the structure can be interpreted as *I*4/*mmm* with two octahedral tilts about the [011] and [01-1] axes. These tilts are highlighted in **Figure 1d** with orange and red arrows indicating different directions of octahedral tilts, while an asterisk marks the consistent octahedral across different views. Transitioning to the *Fmmm* space group, the octahedral tilts are absent, resulting in a structure that might otherwise resemble the *I*4/*mmm* symmetry. However, as shown in **Figure 1e**, lattice distortions along the *a* and *b* axes prevent this. Furthermore, it is worth mentioning that the *I*4/*mmm* structure does not display a perfect square lattice. This difference is primarily due to a slight displacement of oxygen along the *c*-axis, as represented in **Figure 1f**, which disrupts the planarity between the oxygen and nickel atoms, inducing nonlinear "in-plane" Ni–O–Ni bond angles.

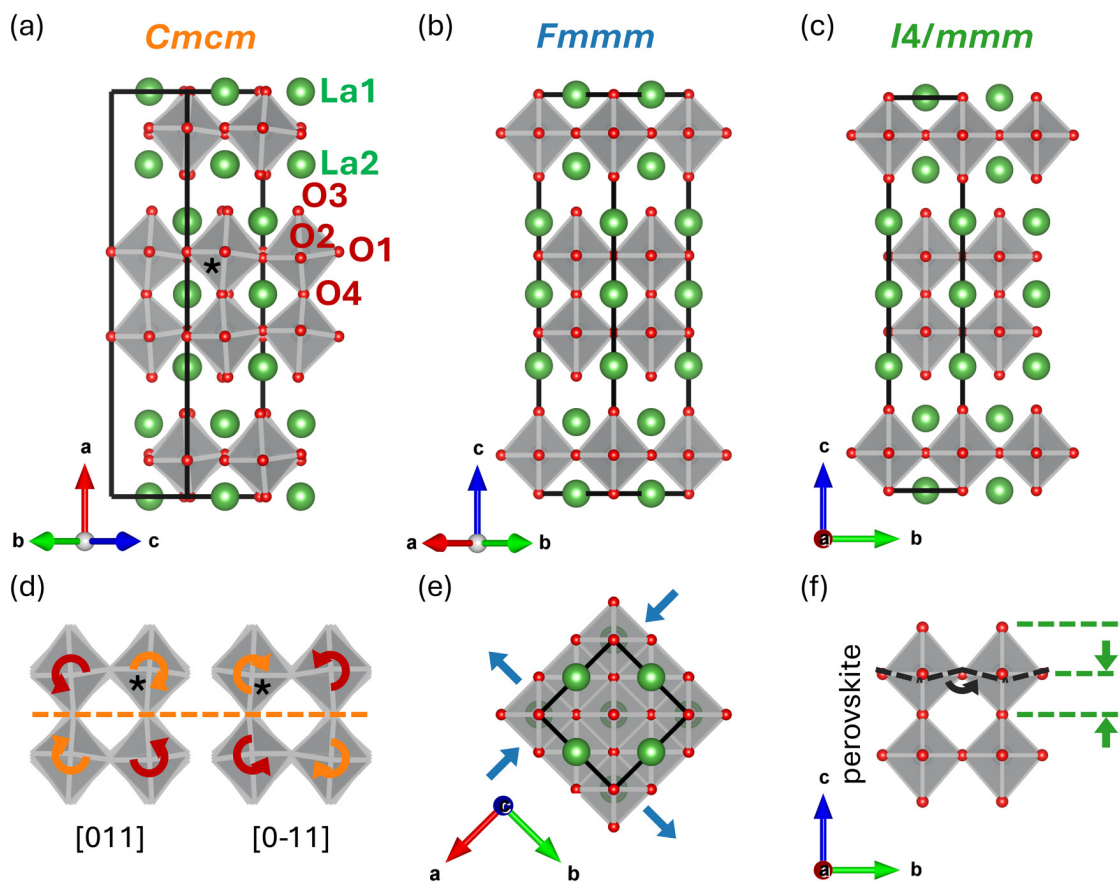

**Figure 1.** Crystal structure of LNO-2222. **(a-c)** Layer stacking views in the space group *Cmcm*, *Fmmm*, and *I4/mmm*, respectively. Green, grey, and red represent La, Ni, and O atoms. In the *Cmcm* space group, crystallographically unique atoms are labeled. **(d)** Two octahedral tilts about the [011] and [0-11] axes in the *Cmcm* space group. **(e)** Lattice distortion along the *a* and *b* axes in the *Fmmm* space group. **(f)** Oxygen displacements along the *c* axis in the *I4/mmm* space group, leading to nonlinear "in-plane" Ni–O–Ni bond angles.

The crystallographic data and structure refinements for LNO-2222 at 80 K and 280 K are summarized in **Tables 1–3**. Our refinements revealed no instances of oxygen vacancies. A slight lattice expansion was observed at 280 K when compared to the structure at 80 K as expected for thermal expansion. Our experimental reciprocal lattice planes, (0*kl*), (*h*1*l*), and (*hk*0) at temperatures of 80 K, 280 K, and 400 K, are presented in **Figure 2**. To enable a clearer visual comparison, the *Cmcm* unit cell was transformed into a symmetry-equivalent *Amam* setting, with the *c* axis oriented perpendicular to the perovskite layers. Laue symmetry *mmm* was applied in the regeneration of these (*hkl*) planes. The reflection conditions for the *Amam* space group are defined as $k + l = 2n$ for all reflections, and $h = 2n$ specifically for reflections on the (*h*0*l*) plane. For the *F*-centering lattice, the reflection conditions stipulate that *h*, *k*, *l* must all be either even or odd ('unmixed'). In our analysis, additional reflections that characterize the *Amam* space group and signify violations for the *Fmmm* space group were observed and labeled on each reciprocal lattice plane. These reflections progressively become weaker as temperatures increased. Further details on the crystallographic data, structure refinements, and reciprocal lattice planes at other temperatures studied are available in **Figures S1**, **S2**, and **Tables S1–S14**.

**Table 1.** Crystal data and structure refinement of LNO-2222 at 80 K and 280 K.

| Chemical formula | La$_3$Ni$_2$O$_7$-2222 | La$_3$Ni$_2$O$_7$-2222 |
|---|---|---|
| Temperature | 80(2) K | 280(2) K |
| Formula weight | 646.15 g/mol | 646.15 g/mol |
| Space group | *Cmcm* | *Cmcm* |
| Unit cell dimensions | $a$ = 20.4840(6) Å | $a$ = 20.5290(6) Å |
|  | $b$ = 5.44800(16) Å | $b$ = 5.44684(18) Å |
|  | $c$ = 5.37711(17) Å | $c$ = 5.38990(18) Å |
| Volume | 600.07(3) Å$^3$ | 602.69(3) Å$^3$ |
| Z | 4 | 4 |
| Density (calculated) | 7.152 g/cm$^3$ | 7.121 g/cm$^3$ |
| Absorption coefficient | 27.042 mm$^{-1}$ | 26.924 mm$^{-1}$ |
| $F(000)$ | 1132 | 1132 |
| $\theta$ range | 3.87 to 40.76° | 3.87 to 40.71° |
| Reflections collected | 18645 | 18646 |
| Independent reflections | 1054 [$R_{int}$ = 0.0682] | 1056 [$R_{int}$ = 0.0663] |
| Refinement method | Full-matrix least-squares on $F^2$ | Full-matrix least-squares on $F^2$ |
| Data / restraints / parameters | 1054 / 0 / 37 | 1056 / 0 / 37 |
| Final $R$ indices | $R_1$ ($I$>2$\sigma$($I$)) = 0.0280; $wR_2$ ($I$>2$\sigma$($I$)) = 0.0695 | $R_1$ ($I$>2$\sigma$($I$)) = 0.0265; $wR_2$ ($I$>2$\sigma$($I$)) = 0.0670 |
|  | $R_1$ (all) = 0.0341; $wR_2$ (all) = 0.0720 | $R_1$ (all) = 0.0358; $wR_2$ (all) = 0.0711 |
| Largest diff. peak and hole | +5.390 e/Å$^{-3}$ and −1.962 e/Å$^{-3}$ | +6.066 e/Å$^{-3}$ and −1.831 e/Å$^{-3}$ |
| R.M.S. deviation from mean | 0.461 e/Å$^{-3}$ | 0.439 e/Å$^{-3}$ |
| Goodness-of-fit on $F^2$ | 1.184 | 1.109 |

**Table 2.** Atomic coordinates and equivalent isotropic atomic displacement parameters (Å$^2$) of LNO-2222 at 80 K. $U_{eq}$ is defined as one third of the trace of the orthogonalized $U_{ij}$ tensor.

|     | Wyck. | x           | y           | z   | Occ. | $U_{eq}$     |
|-----|-------|-------------|-------------|-----|------|--------------|
| La$_1$ | 4c    | 0           | 0.75089(5)  | 1/4 | 1    | 0.00354(7)   |
| La$_2$ | 8g    | 0.32024(2)  | 0.25870(3)  | 1/4 | 1    | 0.00340(7)   |
| Ni    | 8g    | 0.09589(3)  | 0.25256(8)  | 1/4 | 1    | 0.00287(10)  |
| O$_1$ | 8e    | 0.39468(16) | 0           | 0   | 1    | 0.0062(5)    |
| O$_2$ | 8e    | 0.08878(15) | 0           | 0   | 1    | 0.0063(5)    |
| O$_3$ | 8g    | 0.20448(16) | 0.2140(6)   | 1/4 | 1    | 0.0080(5)    |
| O$_4$ | 4c    | 0           | 0.2944(9)   | 1/4 | 1    | 0.0066(6)    |

**Table 3.** Atomic coordinates and equivalent isotropic atomic displacement parameters (Å$^2$) of LNO-2222 at 280 K. $U_{eq}$ is defined as one third of the trace of the orthogonalized $U_{ij}$ tensor.

|     | Wyck. | x           | y           | z   | Occ. | $U_{eq}$     |
|-----|-------|-------------|-------------|-----|------|--------------|
| La$_1$ | 4c    | 0           | 0.75058(4)  | 1/4 | 1    | 0.00729(7)   |
| La$_2$ | 8g    | 0.32019(2)  | 0.25789(3)  | 1/4 | 1    | 0.00634(7)   |
| Ni    | 8g    | 0.09589(3)  | 0.25242(7)  | 1/4 | 1    | 0.00472(9)   |
| O$_1$ | 8e    | 0.39534(16) | 0           | 0   | 1    | 0.0115(5)    |
| O$_2$ | 8e    | 0.08932(15) | 0           | 0   | 1    | 0.0109(5)    |
| O$_3$ | 8g    | 0.20441(16) | 0.2173(6)   | 1/4 | 1    | 0.0129(5)    |
| O$_4$ | 4c    | 0           | 0.2909(8)   | 1/4 | 1    | 0.0117(7)    |

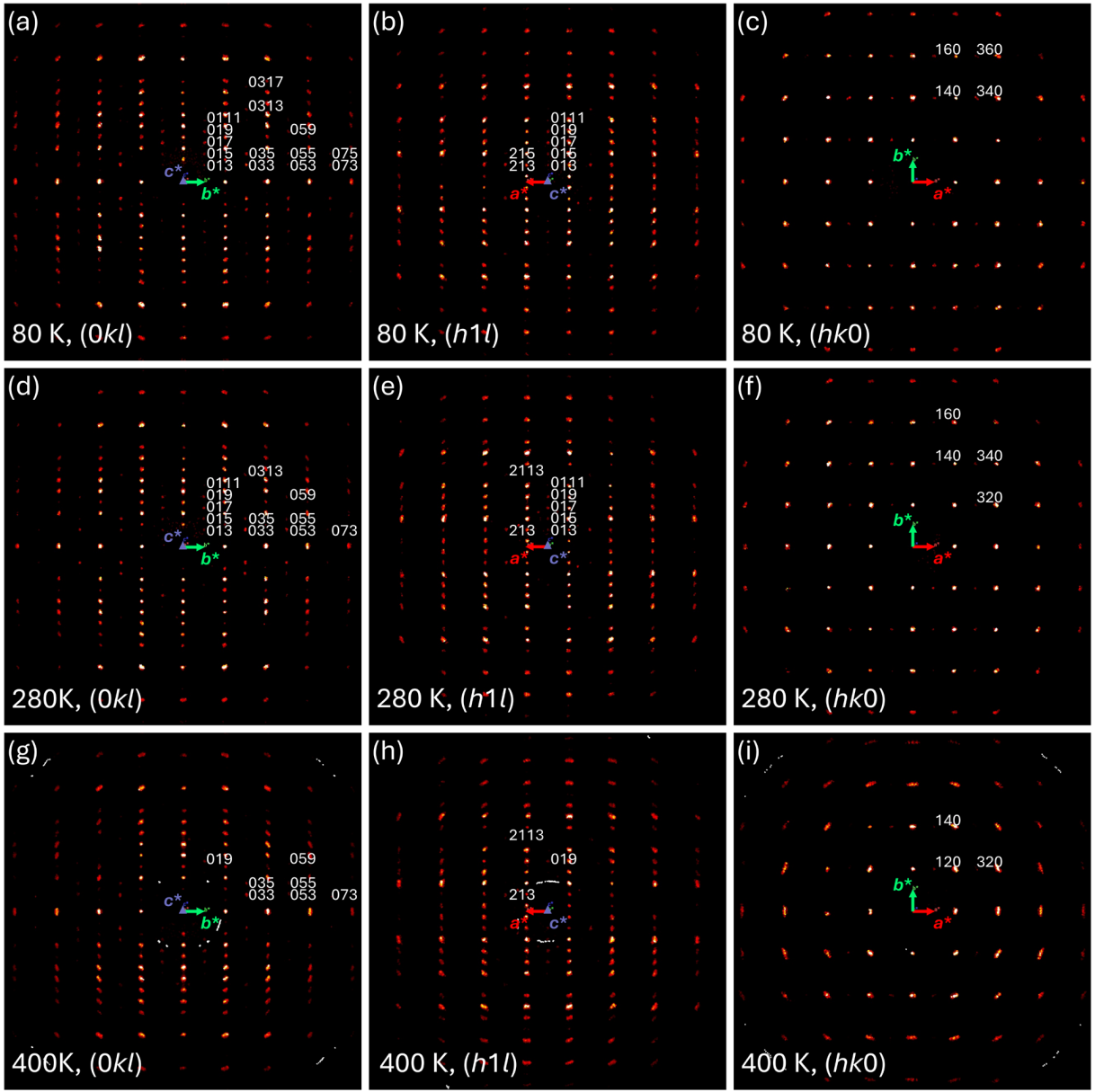

**Figure 2.** Reciprocal lattice planes of LNO-2222. **(a–c)** (0*kl*), (*h*1*l*) and (*hk*0) planes at 80 K. **(d–f)** (0*kl*), (*h*1*l*) and (*hk*0) planes at 280 K. **(g–i)** (0*kl*), (*h*1*l*) and (*hk*0) planes at 400 K. The *Cmcm* unit cell has been transformed into the *Amam* setting to enhance visual clarity. Laue symmetry *mmm* has been applied in the regeneration of these (*hkl*) planes.

**Figure 3a** and **3b** display the temperature-dependent evolution of the lattice parameters, *a*, *b*, *c*, and the unit cell volume of LNO-2222. These parameters follow the expected thermal expansion behavior as temperature increases. Notably, the evolution of the lattice parameter *b* is non-monotonic, exhibiting a distinct anomaly at 120 K. One possible explanation is structure modulations, potentially through charge ordering or the formation of density waves. **Figure 3c** and **3d** provide more structure details about the in-plane and out-of-plane Ni–O–Ni bond angles. As temperature decreases, these bond angles increasingly deviate from 180 degrees, indicative of enhanced octahedral tilts. Such behavior points to potential signs of symmetry lowering. Outlined within **Scheme 1** is the methodology for analyzing experimental reflection data. We identify and focus on specific reflections characteristic of *C*-centering and violate *F*-centering norms, which are expected to be weak, as well as reflections common to both *C*- and *F*-centering. Due to data redundancy, the same reflections are observed multiple times during data collection. We also merge reflections equivalent under Laue symmetry to streamline the analysis. Moreover, we define the average intensity ratio of specific reflections from these two classes, for example, 041 and 002 here, as a metric to evaluate potential changes in symmetry. The unbiased selection of reflections has been validated by incorporating an additional *C*-centering characteristic reflection, 330, which is among the most intense. The results are presented in **Figure S3**, demonstrating the consistency and reliability of our methodology. Our comprehensive analysis confirms that at low temperatures, down to 80 K, the *Cmcm* space group becomes increasingly favorable, aligning with the observed changes in the out-of-plane Ni–O4–Ni bond angle as detailed in **Figure 3c**. This correlation emphasizes the structural dynamics of LNO-2222 under varying thermal conditions.

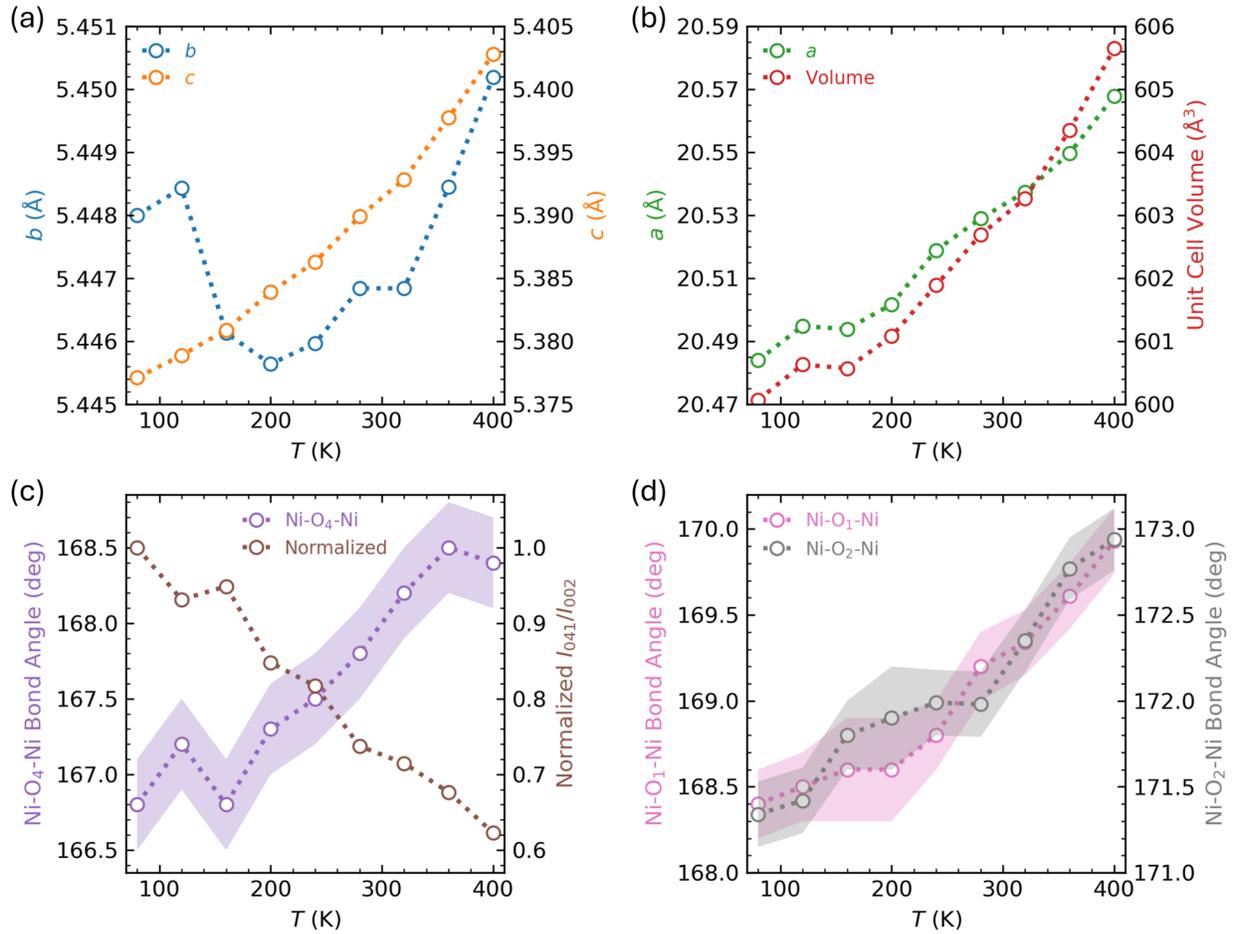

**Figure 3.** Temperature-dependent structure evolution of LNO-2222. **(a)** Lattice parameters $b$ and $c$. **(b)** Lattice parameter $a$ and unit cell volume. **(c)** Out-of-plane Ni–O4–Ni bond angle and the observed intensity ratio of the 041 and 002 reflections, normalized to 80 K. **(d)** In-plane Ni–O1–Ni and Ni–O2–Ni bond angles. Lines serve as guides to the eye, and error bars are indicated by color filling.

Given the observed enhancement of *Cmcm* reflections at low temperatures, considering external pressure as a tuning parameter offers a new perspective on the structure behavior of LNO-2222. We can extrapolate its temperature behavior at ambient pressure to at least slightly higher pressures, under the assumption that the *Cmcm-Fmmm* transition at low temperatures and high pressures is gradual rather than abrupt. This assumption is reasonable, as no formation or breaking of chemical bonds is involved. In terms of crystal structure dynamics, for the transition to occur, the out-of-plane Ni–O–Ni bond angle must approach exactly 180 degrees. Based on this requirement, we hypothesize that, compared to room temperature, a higher pressure would be necessary to induce this transition at lower temperatures. **Figure 4** presents our structure phase diagram of LNO-2222, where the *Cmcm-Fmmm* phase boundary is indicated by a dashed line. This boundary is notably inconsistent with the reported one.[11] Additionally, it is plausible that sufficiently high temperatures could also facilitate this phase transition, although such conditions may extend beyond the scope of our current discussion.

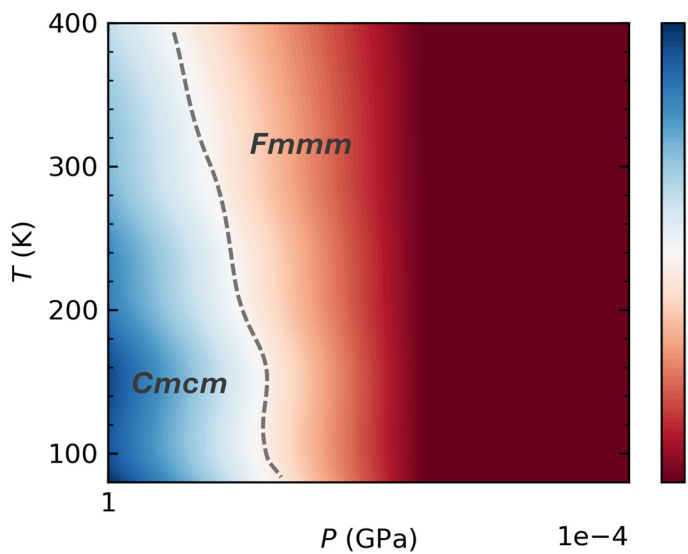

**Figure 4.** Structure phase diagram of LNO-2222. The dashed line indicates the phase boundary between the *Cmcm* and *Fmmm* space groups. The phase diagram extrapolates the ambient pressure behavior of LNO-2222 to a little higher pressure (less than 0.2 GPa). Further experiments under high pressure are necessary for confirmation.

The evaluation of the *Fmmm-I4/mmm* phase boundary, as reported in the referenced study,[11] presents distinct challenges due to inconsistencies in unit cell selection, necessitating the use of √2 super reciprocal vectors in the *hk* plane for accurate analysis. Furthermore, the detection of very weak difference reflections, which appear on half-integral reciprocal lattice planes when using original sub cell axes, requires specialized efforts. These reflections are critical for confirming the phase transition but may not be readily observable under high-pressure conditions due to the inherent experimental challenges. Further detailed experiments involving (low-temperature) high-pressure single crystal XRD will be essential to fully elucidate the complex crystallographic behavior of LNO-2222.

In conclusion, we present our investigation into the temperature-dependent structural evolution of LNO-2222 single crystals at ambient pressure. Our results highlight the enhancement of *Cmcm* reflections as temperature decreases. Additionally, we have developed our structure phase diagram for LNO-2222 that differs from the reported, particularly concerning the *Cmcm-Fmmm* phase boundary. This study not only provides a benchmark method for conducting single crystal structure analysis of LNO-2222 using laboratory X-rays but also delivers high-quality crystallographic data across various temperatures. More importantly, our work enhances the understanding of the complex crystallographic behavior of this system, laying a solid foundation for further experimental and theoretical investigations.

## Supporting Information

Reciprocal lattice planes of LNO-2222 at temperatures of 120 K, 160 K, 200 K, 240 K, 320 K, and 360 K; Validation of unbiased reflection selections; Crystal data and structure refinement, atomic coordinates, and equivalent isotropic atomic displacement parameters of LNO-2222 at temperatures of 120 K, 160 K, 200 K, 240 K, 320 K, 360 K, and 400 K.

## Acknowledgments

The work at Michigan State University was supported by U.S. DOE-BES under Contract DE-SC0023648. The work at the University of Tennessee (crystal growth) was supported by the Air Force Office of Scientific Research under Grant No. FA9550-23-1-0502. H.W. appreciate helpful discussions with Dr. Xinglong Chen (Argonne National Laboratory).

**For Table of Contents Only**

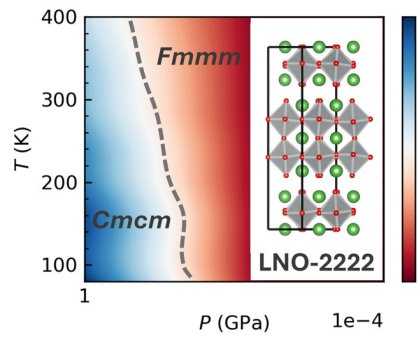

# Supporting Information

# Temperature-dependent Structural Evolution of Ruddlesden–Popper Bilayer Nickelate La$_3$Ni$_2$O$_7$

Haozhe Wang[1], Haidong Zhou[2], Weiwei Xie[1]*

1. Department of Chemistry, Michigan State University, East Lansing, MI, 48824, USA
2. Department of Physics and Astronomy, University of Tennessee, Knoxville, TN, 37996, USA

* Email: xieweiwe@msu.edu

# Table of Contents

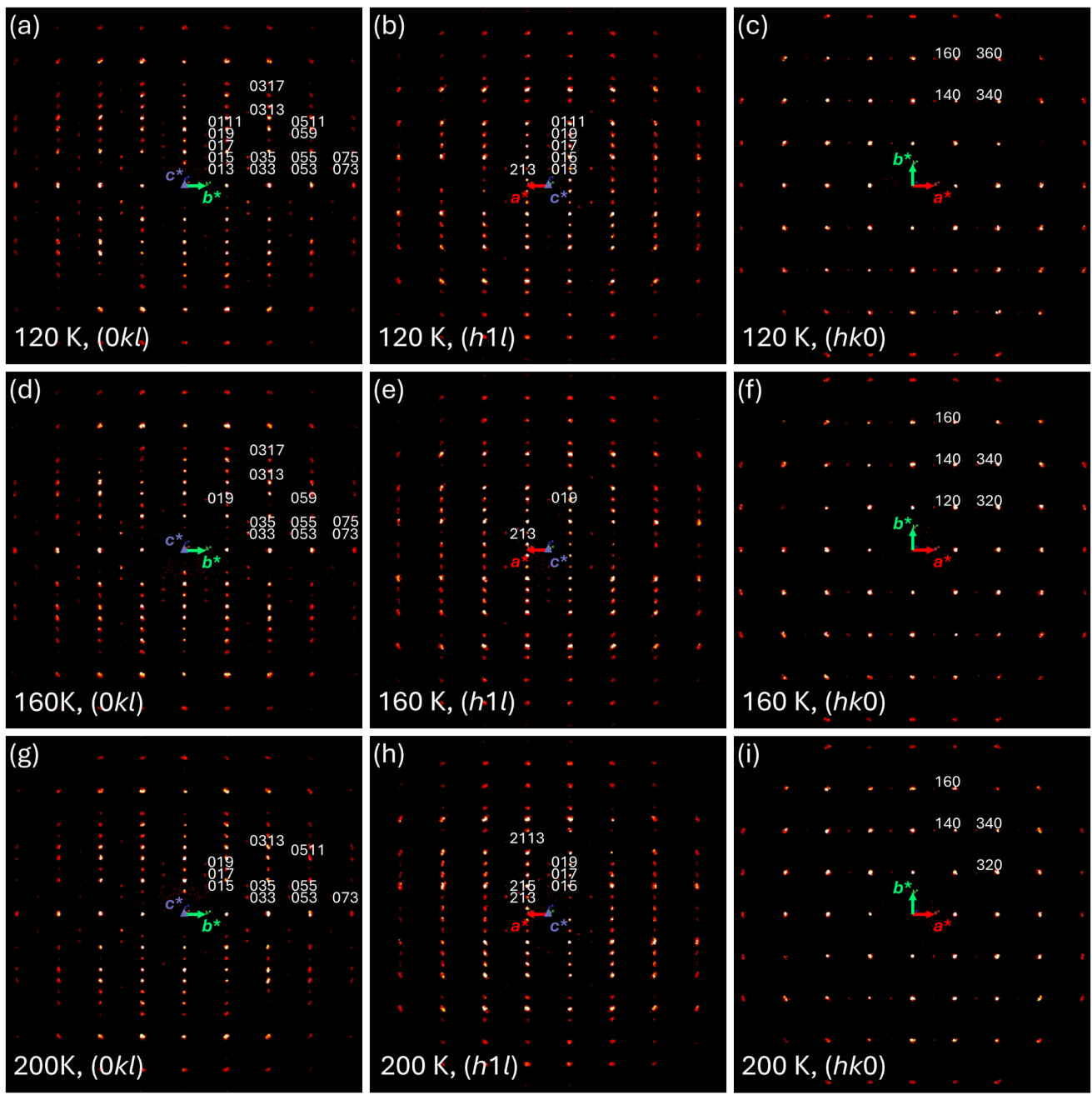

**Figure S1.** Reciprocal lattice planes of LNO-2222. **(a–c)** (0*kl*), (*h*1*l*) and (*hk*0) planes at 120 K. **(d–f)** (0*kl*), (*h*1*l*) and (*hk*0) planes at 160 K. **(g–i)** (0*kl*), (*h*1*l*) and (*hk*0) planes at 200 K. The *Cmcm* unit cell has been transformed into the *Amam* setting to enhance visual clarity. Laue symmetry *mmm* has been applied in the regeneration of these (*hkl*) planes.

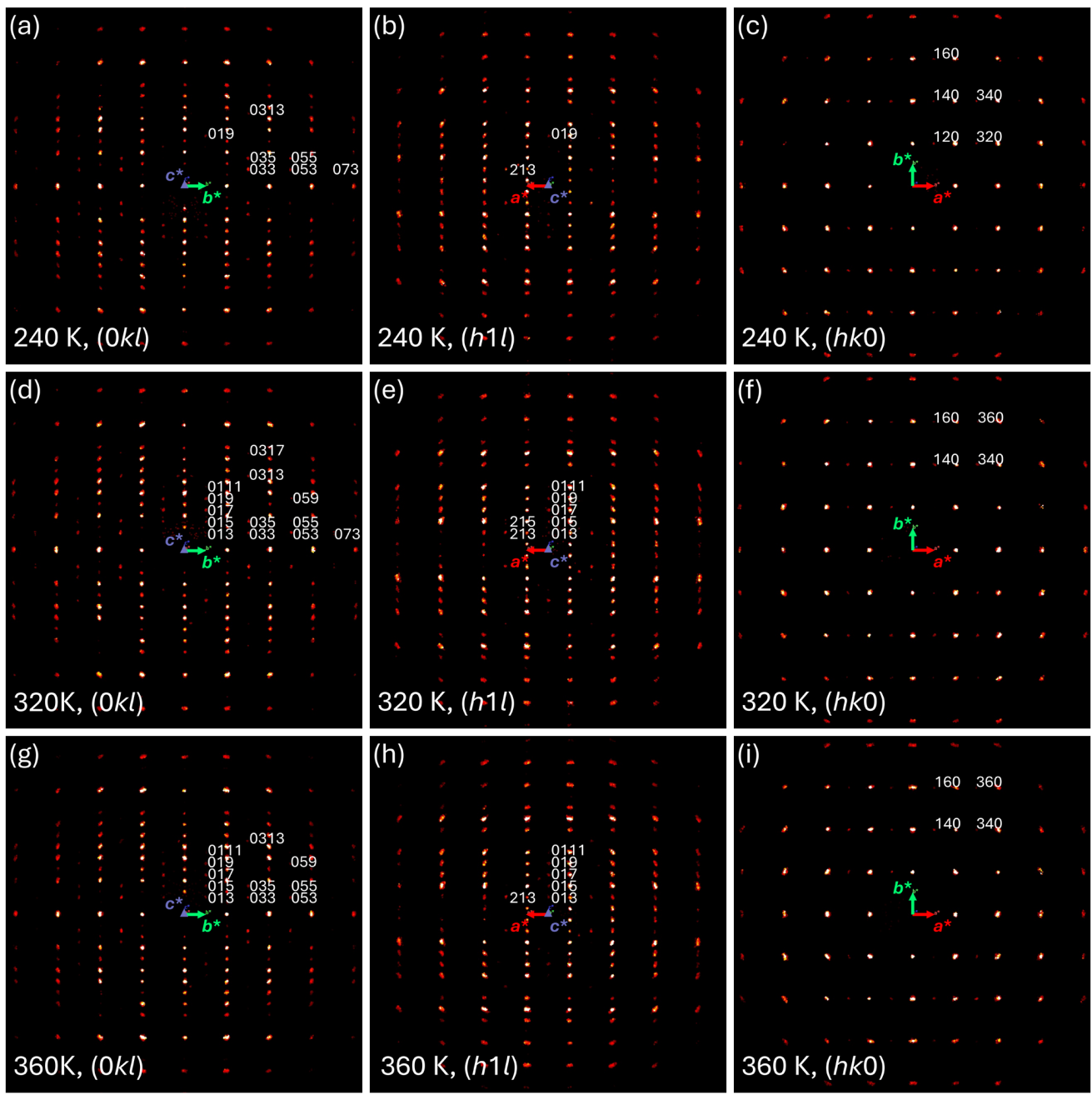

**Figure S2.** Reciprocal lattice planes of LNO-2222. **(a–c)** (0*kl*), (*h*1*l*) and (*hk*0) planes at 240 K. **(d–f)** (0*kl*), (*h*1*l*) and (*hk*0) planes at 320 K. **(g–i)** (0*kl*), (*h*1*l*) and (*hk*0) planes at 360 K. The *Cmcm* unit cell has been transformed to the *Amam* setting to enhance visual clarity. Laue symmetry *mmm* has been applied in the regeneration of these (*hkl*) planes.

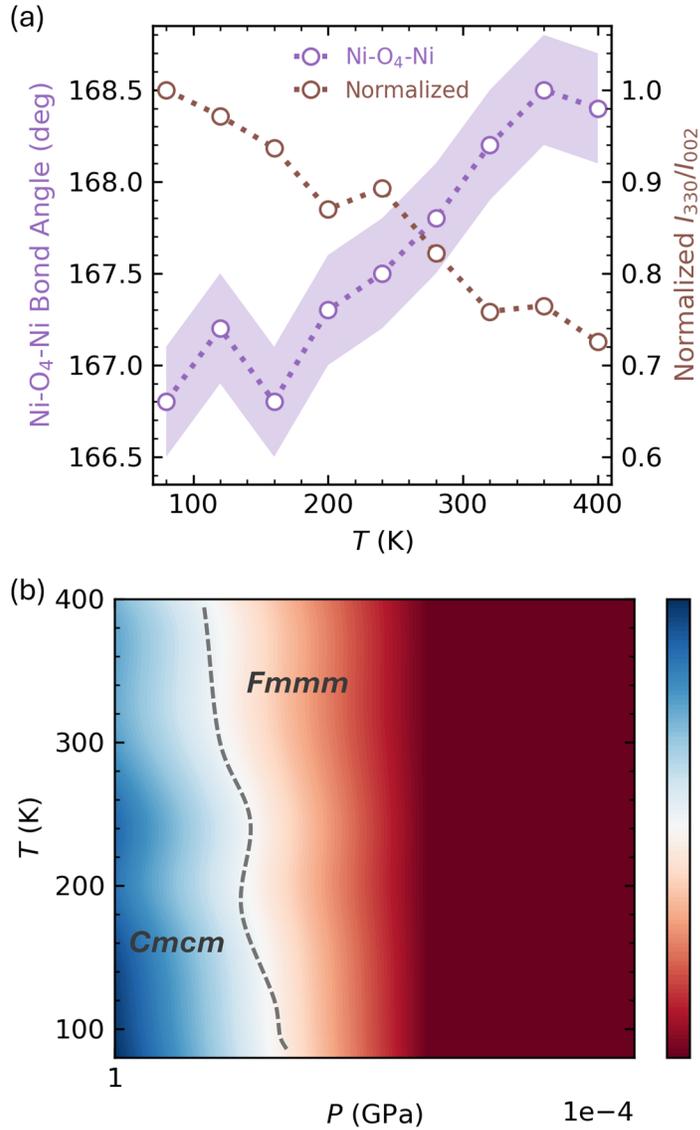

**Figure S3.** Validation of unbiased reflection selections. **(a)** Temperature-dependent evolution of out-of-plane Ni–O4–Ni bond angle and the observed intensity ratio of the 330 and 002 reflections, normalized at 80 K. **(b)** Structure phase diagram of LNO-2222. The dashed line indicates the phase boundary between the *Cmcm* and *Fmmm* space groups.

**Table S1.** Crystal data and structure refinement of LNO-2222 at 120 K.

| Chemical formula | La$_3$Ni$_2$O$_7$-2222 |
|---|---|
| Temperature | 120(2) K |
| Formula weight | 646.15 g/mol |
| Space group | *Cmcm* |
| Unit cell dimensions | $a$ = 20.4948(6) Å |
| | $b$ = 5.44843(15) Å |
| | $c$ = 5.37889(16) Å |
| Volume | 600.63(3) Å$^3$ |
| *Z* | 4 |
| Density (calculated) | 7.146 g/cm$^3$ |
| Absorption coefficient | 27.017 mm$^{-1}$ |
| *F*(000) | 1132 |
| *θ* range | 3.87 to 40.71° |
| Reflections collected | 18774 |
| Independent reflections | 1053 [$R_{int}$ = 0.0635] |
| Refinement method | Full-matrix least-squares on $F^2$ |
| Data / restraints / parameters | 1053 / 0 / 37 |
| Final *R* indices | $R_1$ ($I$>2σ($I$)) = 0.0275; $wR_2$ ($I$>2σ($I$)) = 0.0683 |
| | $R_1$ (all) = 0.0335; $wR_2$ (all) = 0.0710 |
| Largest diff. peak and hole | +6.281 e/Å$^{-3}$ and −1.674 e/Å$^{-3}$ |
| R.M.S. deviation from mean | 0.462 e/Å$^{-3}$ |
| Goodness-of-fit on F$^2$ | 1.165 |

**Table S2.** Atomic coordinates and equivalent isotropic atomic displacement parameters (Å$^2$) of LNO-2222 at 120 K. $U_{eq}$ is defined as one third of the trace of the orthogonalized $U_{ij}$ tensor.

| | Wyck. | *x* | *y* | *z* | Occ. | $U_{eq}$ |
|---|---|---|---|---|---|---|
| **La$_1$** | 4*c* | 0 | 0.75081(5) | 1/4 | 1 | 0.00438(7) |
| **La$_2$** | 8*g* | 0.32024(2) | 0.25860(3) | 1/4 | 1 | 0.00406(7) |
| **Ni** | 8*g* | 0.09591(3) | 0.25254(7) | 1/4 | 1 | 0.00329(10) |
| **O$_1$** | 8*e* | 0.39473(16) | 0 | 0 | 1 | 0.0075(5) |
| **O$_2$** | 8*e* | 0.08887(15) | 0 | 0 | 1 | 0.0072(5) |
| **O$_3$** | 8*g* | 0.20437(16) | 0.2143(6) | 1/4 | 1 | 0.0081(5) |
| **O$_4$** | 4*c* | 0 | 0.2929(9) | 1/4 | 1 | 0.0085(7) |

**Table S3.** Crystal data and structure refinement of LNO-2222 at 160 K.

| Chemical formula | La$_3$Ni$_2$O$_7$-2222 |
|---|---|
| Temperature | 160(2) K |
| Formula weight | 646.15 g/mol |
| Space group | *Cmcm* |
| Unit cell dimensions | $a$ = 20.4939(6) Å |
| | $b$ = 5.44613(17) Å |
| | $c$ = 5.38089(18) Å |
| Volume | 600.57(3) Å$^3$ |
| Z | 4 |
| Density (calculated) | 7.146 g/cm$^3$ |
| Absorption coefficient | 27.019 mm$^{-1}$ |
| F(000) | 1132 |
| $\theta$ range | 3.87 to 40.70° |
| Reflections collected | 18729 |
| Independent reflections | 1053 [$R_{int}$ = 0.0709] |
| Refinement method | Full-matrix least-squares on $F^2$ |
| Data / restraints / parameters | 1053 / 0 / 37 |
| Final $R$ indices | $R_1$ ($I$>2$\sigma$($I$)) = 0.0286; $wR_2$ ($I$>2$\sigma$($I$)) = 0.0684 |
| | $R_1$ (all) = 0.0368; $wR_2$ (all) = 0.0716 |
| Largest diff. peak and hole | +5.998 e/Å$^{-3}$ and –2.019 e/Å$^{-3}$ |
| R.M.S. deviation from mean | 0.468 e/Å$^{-3}$ |
| Goodness-of-fit on F$^2$ | 1.104 |

**Table S4.** Atomic coordinates and equivalent isotropic atomic displacement parameters (Å$^2$) of LNO-2222 at 160 K. $U_{eq}$ is defined as one third of the trace of the orthogonalized $U_{ij}$ tensor.

| | Wyck. | x | y | z | Occ. | $U_{eq}$ |
|---|---|---|---|---|---|---|
| **La$_1$** | 4c | 0 | 0.75085(5) | 1/4 | 1 | 0.00498(7) |
| **La$_2$** | 8g | 0.32023(2) | 0.25840(3) | 1/4 | 1 | 0.00453(7) |
| **Ni** | 8g | 0.09591(3) | 0.25242(8) | 1/4 | 1 | 0.00363(10) |
| **O$_1$** | 8e | 0.39481(17) | 0 | 0 | 1 | 0.0083(5) |
| **O$_2$** | 8e | 0.08915(16) | 0 | 0 | 1 | 0.0076(5) |
| **O$_3$** | 8g | 0.20437(17) | 0.2152(6) | 1/4 | 1 | 0.0091(5) |
| **O$_4$** | 4c | 0 | 0.2942(9) | 1/4 | 1 | 0.0082(7) |

**Table S5.** Crystal data and structure refinement of LNO-2222 at 200 K.

| Chemical formula | La$_3$Ni$_2$O$_7$-2222 |
|---|---|
| Temperature | 200(2) K |
| Formula weight | 646.15 g/mol |
| Space group | *Cmcm* |
| Unit cell dimensions | *a* = 20.5017(7) Å |
| | *b* = 5.44564(18) Å |
| | *c* = 5.38391(19) Å |
| Volume | 601.08(3) Å$^3$ |
| *Z* | 4 |
| Density (calculated) | 7.140 g/cm$^3$ |
| Absorption coefficient | 26.996 mm$^{-1}$ |
| *F*(000) | 1132 |
| $\theta$ range | 3.87 to 40.75° |
| Reflections collected | 18613 |
| Independent reflections | 1057 [$R_{int}$ = 0.0741] |
| Refinement method | Full-matrix least-squares on $F^2$ |
| Data / restraints / parameters | 1057 / 0 / 37 |
| Final *R* indices | $R_1$ (*I*>2$\sigma$(*I*)) = 0.0289; $wR_2$ (*I*>2$\sigma$(*I*)) = 0.0697 |
| | $R_1$ (all) = 0.0384; $wR_2$ (all) = 0.0739 |
| Largest diff. peak and hole | +6.311 e/Å$^{-3}$ and –2.267 e/Å$^{-3}$ |
| R.M.S. deviation from mean | 0.488 e/Å$^{-3}$ |
| Goodness-of-fit on F$^2$ | 1.103 |

**Table S6.** Atomic coordinates and equivalent isotropic atomic displacement parameters (Å$^2$) of LNO-2222 at 200 K. $U_{eq}$ is defined as one third of the trace of the orthogonalized $U_{ij}$ tensor.

| | Wyck. | *x* | *y* | *z* | Occ. | $U_{eq}$ |
|---|---|---|---|---|---|---|
| **La$_1$** | 4*c* | 0 | 0.75074(5) | 1/4 | 1 | 0.00575(8) |
| **La$_2$** | 8*g* | 0.32021(2) | 0.25822(3) | 1/4 | 1 | 0.00517(7) |
| **Ni** | 8*g* | 0.09589(3) | 0.25243(8) | 1/4 | 1 | 0.00393(10) |
| **O$_1$** | 8*e* | 0.39482(18) | 0 | 0 | 1 | 0.0089(6) |
| **O$_2$** | 8*e* | 0.08923(17) | 0 | 0 | 1 | 0.0093(6) |
| **O$_3$** | 8*g* | 0.20430(17) | 0.2155(6) | 1/4 | 1 | 0.0099(6) |
| **O$_4$** | 4*c* | 0 | 0.2926(9) | 1/4 | 1 | 0.0089(7) |

**Table S7.** Crystal data and structure refinement of LNO-2222 at 240 K.

| Chemical formula | La$_3$Ni$_2$O$_7$-2222 |
|---|---|
| Temperature | 240(2) K |
| Formula weight | 646.15 g/mol |
| Space group | Cmcm |
| Unit cell dimensions | a = 20.5189(6) Å |
| | b = 5.44597(17) Å |
| | c = 5.38630(18) Å |
| Volume | 601.89(3) Å$^3$ |
| Z | 4 |
| Density (calculated) | 7.131 g/cm$^3$ |
| Absorption coefficient | 26.960 mm$^{-1}$ |
| F(000) | 1132 |
| θ range | 3.87 to 40.73° |
| Reflections collected | 18656 |
| Independent reflections | 1055 [$R_{int}$ = 0.0649] |
| Refinement method | Full-matrix least-squares on $F^2$ |
| Data / restraints / parameters | 1055 / 0 / 37 |
| Final R indices | $R_1$ ($I>2\sigma(I)$) = 0.0271; $wR_2$ ($I>2\sigma(I)$) = 0.0672 |
| | $R_1$ (all) = 0.0360; $wR_2$ (all) = 0.0711 |
| Largest diff. peak and hole | +6.097 e/Å$^{-3}$ and −2.068 e/Å$^{-3}$ |
| R.M.S. deviation from mean | 0.442 e/Å$^{-3}$ |
| Goodness-of-fit on F$^2$ | 1.126 |

**Table S8.** Atomic coordinates and equivalent isotropic atomic displacement parameters (Å$^2$) of LNO-2222 at 240 K. $U_{eq}$ is defined as one third of the trace of the orthogonalized $U_{ij}$ tensor.

|  | Wyck. | x | y | z | Occ. | $U_{eq}$ |
|---|---|---|---|---|---|---|
| La$_1$ | 4c | 0 | 0.75067(5) | 1/4 | 1 | 0.00656(7) |
| La$_2$ | 8g | 0.32020(2) | 0.25804(3) | 1/4 | 1 | 0.00581(7) |
| Ni | 8g | 0.09591(3) | 0.25247(7) | 1/4 | 1 | 0.00435(9) |
| O$_1$ | 8e | 0.39502(16) | 0 | 0 | 1 | 0.0100(5) |
| O$_2$ | 8e | 0.08934(15) | 0 | 0 | 1 | 0.0095(5) |
| O$_3$ | 8g | 0.20454(15) | 0.2156(6) | 1/4 | 1 | 0.0111(5) |
| O$_4$ | 4c | 0 | 0.2922(8) | 1/4 | 1 | 0.0110(7) |

**Table S9.** Crystal data and structure refinement of LNO-2222 at 320 K.

| Chemical formula | La$_3$Ni$_2$O$_7$-2222 |
|---|---|
| Temperature | 320(2) K |
| Formula weight | 646.15 g/mol |
| Space group | *Cmcm* |
| Unit cell dimensions | $a$ = 20.5374(7) Å |
| | $b$ = 5.44684(18) Å |
| | $c$ = 5.39285(19) Å |
| Volume | 603.27(4) Å$^3$ |
| Z | 4 |
| Density (calculated) | 7.114 g/cm$^3$ |
| Absorption coefficient | 26.899 mm$^{-1}$ |
| F(000) | 1132 |
| $\theta$ range | 3.87 to 40.83° |
| Reflections collected | 18690 |
| Independent reflections | 1060 [$R_{int}$ = 0.0591] |
| Refinement method | Full-matrix least-squares on $F^2$ |
| Data / restraints / parameters | 1060 / 0 / 37 |
| Final $R$ indices | $R_1$ ($I$>2$\sigma$($I$)) = 0.0258; $wR_2$ ($I$>2$\sigma$($I$)) = 0.0640 |
| | $R_1$ (all) = 0.0348; $wR_2$ (all) = 0.0677 |
| Largest diff. peak and hole | +6.146 e/Å$^{-3}$ and –1.499 e/Å$^{-3}$ |
| R.M.S. deviation from mean | 0.433 e/Å$^{-3}$ |
| Goodness-of-fit on F$^2$ | 1.114 |

**Table S10.** Atomic coordinates and equivalent isotropic atomic displacement parameters (Å$^2$) of LNO-2222 at 320 K. $U_{eq}$ is defined as one third of the trace of the orthogonalized $U_{ij}$ tensor.

| | Wyck. | $x$ | $y$ | $z$ | Occ. | $U_{eq}$ |
|---|---|---|---|---|---|---|
| **La$_1$** | 4$c$ | 0 | 0.75055(4) | 1/4 | 1 | 0.00810(7) |
| **La$_2$** | 8$g$ | 0.32020(2) | 0.25766(3) | 1/4 | 1 | 0.00700(6) |
| **Ni** | 8$g$ | 0.09589(3) | 0.25237(6) | 1/4 | 1 | 0.00515(9) |
| **O$_1$** | 8$e$ | 0.39545(15) | 0 | 0 | 1 | 0.0129(5) |
| **O$_2$** | 8$e$ | 0.08962(14) | 0 | 0 | 1 | 0.0121(5) |
| **O$_3$** | 8$g$ | 0.20435(15) | 0.2179(5) | 1/4 | 1 | 0.0137(5) |
| **O$_4$** | 4$c$ | 0 | 0.2898(7) | 1/4 | 1 | 0.0126(6) |

**Table S11.** Crystal data and structure refinement of LNO-2222 at 360 K.

| Chemical formula | La$_3$Ni$_2$O$_7$-2222 |
|---|---|
| Temperature | 360(2) K |
| Formula weight | 646.15 g/mol |
| Space group | *Cmcm* |
| Unit cell dimensions | *a* = 20.5497(6) Å |
| | *b* = 5.44845(18) Å |
| | *c* = 5.39774(18) Å |
| Volume | 604.35(3) Å$^3$ |
| *Z* | 4 |
| Density (calculated) | 7.102 g/cm$^3$ |
| Absorption coefficient | 26.850 mm$^{-1}$ |
| *F*(000) | 1132 |
| $\theta$ range | 3.87 to 40.91° |
| Reflections collected | 18283 |
| Independent reflections | 1062 [$R_{int}$ = 0.0566] |
| Refinement method | Full-matrix least-squares on $F^2$ |
| Data / restraints / parameters | 1062 / 0 / 37 |
| Final *R* indices | $R_1$ (*I*>2σ(*I*)) = 0.0258; $wR_2$ (*I*>2σ(*I*)) = 0.0654 |
| | $R_1$ (all) = 0.0333; $wR_2$ (all) = 0.0686 |
| Largest diff. peak and hole | +6.010 e/Å$^{-3}$ and –1.941 e/Å$^{-3}$ |
| R.M.S. deviation from mean | 0.409 e/Å$^{-3}$ |
| Goodness-of-fit on F$^2$ | 1.079 |

**Table S12.** Atomic coordinates and equivalent isotropic atomic displacement parameters (Å$^2$) of LNO-2222 at 360 K. $U_{eq}$ is defined as one third of the trace of the orthogonalized $U_{ij}$ tensor.

| | Wyck. | *x* | *y* | *z* | Occ. | $U_{eq}$ |
|---|---|---|---|---|---|---|
| La$_1$ | 4*c* | 0 | 0.75044(4) | 1/4 | 1 | 0.00881(7) |
| La$_2$ | 8*g* | 0.32020(2) | 0.25748(3) | 1/4 | 1 | 0.00760(6) |
| Ni | 8*g* | 0.09586(3) | 0.25234(6) | 1/4 | 1 | 0.00549(9) |
| O$_1$ | 8*e* | 0.39570(15) | 0 | 0 | 1 | 0.0137(5) |
| O$_2$ | 8*e* | 0.08994(14) | 0 | 0 | 1 | 0.0127(5) |
| O$_3$ | 8*g* | 0.20439(14) | 0.2190(5) | 1/4 | 1 | 0.0143(5) |
| O$_4$ | 4*c* | 0 | 0.2887(8) | 1/4 | 1 | 0.0138(6) |

**Table S13.** Crystal data and structure refinement of LNO-2222 at 400 K.

| Chemical formula | La$_3$Ni$_2$O$_7$-2222 |
|---|---|
| Temperature | 400(2) K |
| Formula weight | 646.15 g/mol |
| Space group | *Cmcm* |
| Unit cell dimensions | $a$ = 20.5679(7) Å |
| | $b$ = 5.45019(19) Å |
| | $c$ = 5.4028(2) Å |
| Volume | 605.65(4) Å$^3$ |
| Z | 4 |
| Density (calculated) | 7.086 g/cm$^3$ |
| Absorption coefficient | 26.793 mm$^{-1}$ |
| $F(000)$ | 1132 |
| $\theta$ range | 3.87 to 40.87° |
| Reflections collected | 18515 |
| Independent reflections | 1062 [$R_{int}$ = 0.0566] |
| Refinement method | Full-matrix least-squares on $F^2$ |
| Data / restraints / parameters | 1062 / 0 / 37 |
| Final $R$ indices | $R_1$ ($I$>2$\sigma$($I$)) = 0.0266; $wR_2$ ($I$>2$\sigma$($I$)) = 0.0678 |
| | $R_1$ (all) = 0.0352; $wR_2$ (all) = 0.0715 |
| Largest diff. peak and hole | +6.303 e/Å$^{-3}$ and –1.854 e/Å$^{-3}$ |
| R.M.S. deviation from mean | 0.427 e/Å$^{-3}$ |
| Goodness-of-fit on F$^2$ | 1.133 |

**Table S14.** Atomic coordinates and equivalent isotropic atomic displacement parameters (Å$^2$) of LNO-2222 at 400 K. $U_{eq}$ is defined as one third of the trace of the orthogonalized $U_{ij}$ tensor.

| | Wyck. | x | y | z | Occ. | $U_{eq}$ |
|---|---|---|---|---|---|---|
| **La$_1$** | 4$c$ | 0 | 0.75044(4) | 1/4 | 1 | 0.00969(7) |
| **La$_2$** | 8$g$ | 0.32020(2) | 0.25748(3) | 1/4 | 1 | 0.00819(7) |
| **Ni** | 8$g$ | 0.09586(3) | 0.25234(6) | 1/4 | 1 | 0.00592(9) |
| **O$_1$** | 8$e$ | 0.39570(15) | 0 | 0 | 1 | 0.0150(5) |
| **O$_2$** | 8$e$ | 0.09007(14) | 0 | 0 | 1 | 0.0137(5) |
| **O$_3$** | 8$g$ | 0.20438(14) | 0.2191(5) | 1/4 | 1 | 0.0152(5) |
| **O$_4$** | 4$c$ | 0 | 0.2892(7) | 1/4 | 1 | 0.0141(6) |